\documentclass{scrartcl}
\usepackage{amssymb}
\usepackage{hyperref}
\usepackage{amsmath}
\usepackage{graphicx}
\usepackage{url}

\newcommand{\Tr}{\mathop{\mathrm{Tr}}\nolimits}

\begin{document}

\title{Many-body spin glasses in the microcanonical ensemble \footnote{Dedicated to Professor David Sherrington on the occasion of his 70th birthday.} $^{,}$\footnote{This is a preprint of an article accepted for publication in the Philosophical Magazine 2011 \copyright Taylor
\& Francis; Philosophical Magazine is available online at:\newline
\url{http://www.informaworld.com/smpp/title \~ db=al \~ content=t713695589}
} }

\author{Zsolt Bertalan\footnote{
Corresponding author. Email: zsolt@stat.phys.titech.ac.jp}$\ $ and Hidetoshi Nishimori}

\maketitle

\begin{center}

Department of Physics, Tokyo Institute of Technology, Oh-Okayama, Meguro-ku,
Tokyo 152-8551, Japan

\end{center}

\begin{abstract}
We investigate the $p$-spin model with Gaussian-distributed random interactions in the microcanonical ensemble using the replica theory. For $p=2$, there are only second-order phase transitions and we recover the results of Sherrington and Kirkpatrick obtained in the canonical ensemble. For $p\geq 3$, the transition between the ferromagnetic and paramagnetic phases is of first order, and the microcanonical and canonical ensembles give different results. We also discuss the ensemble inequivalence of the random energy model, corresponding to the limit $p\rightarrow\infty$. This is the first systematic treatment of spin glasses with long-range interactions in the microcanonical ensemble, which shows how the two ensembles give different results.
\end{abstract}

\section{Introduction}
Theoretical aspects of spin glasses attract interest not only from the  statistical-physics community, but also from many other areas of science. The methods of spin glass theory have become indispensable tools for fields like statistical mechanics, information processing, image restoration and neural networks, see \cite{nishimori,mezard-montanari} and references therein. Most models discussed in the mean-field context of spin glasses have long-range interactions, a typical case being the infinite-range Sherrington-Kirkpatrick model \cite{SK}. The purpose of the present paper is to investigate whether or not the canonical and microcanonical ensembles give different results on such systems with long-range interactions.

Interactions are said to be long range when they decay like $\sim 1/r^{\alpha}$, where $r$ is the spatial distance between two interacting objects, and $\alpha$ is less than or equal to the dimension of the system, or when the range of the interactions is of the same order as the system size \cite{campa}. Systems with long-range interactions appear in virtually all areas of physics ranging from atomic physics to gravitational systems (see \cite{campa,dauxois,mukamel2,dauxois2,chavanis} for review). Recent results on long-range interacting systems  \cite{campa,hertel,barre,leyvraz,barre2,mukamel,mukamel2,mukamel3,dauxois,dauxois2,chavanis,bkmn} seem to indicate that we have to rethink the long-held principle that the physical properties of a system are independent of the ensemble they were derived from, especially where first order transitions are concerned \cite{ellis,ellis2,bouchet}. Related phenomena have been observed in driven systems with local dynamics \cite{leder1,leder2}. The main reason for ensemble inequivalence in systems with long-range interactions is the lack of additivity of the energy: If two identical systems with energy $E$ are brought together, the total energy is not $2E$ because the interactions between the two subsystems are of the same order of magnitude as $E$. This suggests us to be cautious in the interpretation of the standard derivation of the canonical ensemble from the microcanonical ensemble.

In a typical phase diagram of many materials, with the temperature and an external control parameter as axes, first-order phase boundaries are simple lines in the canonical ensemble, but in the microcanonical ensemble the question is sometimes more involved because there the temperature is not the principal control parameter. The temperature has to be derived from the equilibrium entropy, and this fact can sometimes have puzzling consequences, including negative specific heat, for systems with long-range interactions \cite{lyb-wood,lyndenbell,thirring}. Experimentally, negative specific heat has been observed for a small cluster of atoms \cite{schmidt}.

Another feature of ensemble inequivalence is the appearance of a mixed phase instead of a clear phase separation in the region of a first-order transition \cite{barre,mukamel,bkmn}. Ergodicity breaking can also occur, when the thermodynamical phase space is not convex. This means that some of the intermediate states between two phase-space points are not realizable, inhibiting a smooth interpolation between these two points. Then, metastable states in a region of the phase space have diverging decay times since the system cannot smoothly evolve toward the true equilibrium state in another region of the phase space. When many-body interactions come into play, the ergodicity can already be broken before the phase transition occurs, inhibiting equilibrium phase transitions \cite{buyl-bouchet,buyl-bouchet2,bkmn}. 

Spin systems on a lattice have been a tool of choice for detailed theoretical studies since many models are tractable analytically and show many features of ensemble inequivalence \cite{barre, mukamel,buyl-bouchet,buyl-bouchet2,bkmn,ispolatov}. A paradigmatic, much studied system with long-range interactions is the infinite-range $p$-spin model, also known as the Ising model with infinite-range $p$-body interactions. It has been studied extensively in the canonical ensemble in the context of spin glasses. For, $p=2$ the model is known as the Sherrington-Kirkpatrick model \cite{SK}. For finite $p$, the $p$-spin model has been discussed in many papers, two of the most recent ones being \cite{krzde,krzde2}, while in the limit $p\rightarrow\infty$ it is known as the random energy model \cite{derrida}. In a recent paper \cite{bkmn} we investigated the ferromagnetic $p$-spin model in random fields and have shown that the canonical and microcanonical ensembles lead to completely different phase diagrams for $p\geq 3$. To discuss ensemble inequivalence in systems with randomness in the interactions, including spin glasses, however, new tools are required. In this paper we extend the replica trick to the microcanonical ensemble and apply it to the $p$-spin model with Gaussian-distributed random bonds and study whether the two ensembles lead to the same or different consequences. 

The paper is organized as follows. In section \ref{sec:model} we introduce the model and sketch the derivation of the replicated microcanonical entropy, with detailed calculations being given in the Appendix \ref{derivation}. We compare the canonical and microcanonical phase diagrams for finite $p$ and take the limit $p\rightarrow\infty$ in section \ref{sec:comp}. We conclude with a discussion and summary in section \ref{sec:summary}.  

\section{Model and entropy}\label{sec:model}
In this section we introduce the model and explain the derivation of the microcanonical entropy and show its replica symmetric form.

The $p$-spin model consists of $N$ spins (in our case Ising spins $S_i=\pm 1$, $i=1,2,\ldots,N$), where each spin interacts with all other spins through $p$-body interactions. Its Hamiltonian is given by
\begin{equation}\label{HAM}
H=-\sum_{i_1<\ldots <i_p} J_{i_1,\ldots ,i_p} S_{i_1}\ldots S_{i_p}.
\end{equation}
In this paper the bonds (interactions) are quenched random numbers obeying a Gaussian distribution
\begin{equation}\label{PJij}
P(J_{i_1...i_p})=\left(\dfrac{N^{p-1}}{\pi J^2 p!}\right)^{1/2}\exp \left(-{\left(J_{i_1...i_p}-\dfrac{j_0p!}{N^{p-1}}\right)^2\dfrac{N^{p-1}}{J^2 p!}}\right),
\end{equation}
with mean $j_0$, called the ferromagnetic bias. For $p=2$, this model is known as the Sherrington-Kirkpatrick model \cite{SK} and has been studied extensively in the canonical ensemble (see e.g. \cite{nishimori}). 

The thermodynamic potential in the microcanonical ensemble is the entropy, which is defined as the logarithm of the sum of states having some given energy, $\mathcal{S}(\epsilon)=\ln \Omega$, where we have chosen the unit $k_{\textrm{B}}=1$. Thus, we have to take the trace of a delta function over the spin configurations,
\begin{equation*}
\Omega=\Tr\delta(N\epsilon-H),
\end{equation*}
where $\epsilon$ is the energy per spin. Since we have to average over the quenched randomness of bonds, equation \eqref{PJij}, we employ the replica trick as detailed in Appendix \ref{derivation}. For the entropy per spin, $s=\mathcal{S}/N$, we get
\begin{equation*}
[\Omega^n]=\exp \left(nNs\right),
\end{equation*}
where we will have to take the limits $N\rightarrow \infty$ and $n\rightarrow 0$ in the end. According to equation \eqref{OMEGA}, the entropy per spin is given by
\begin{eqnarray}\label{entropy}
ns=-\sum_{\alpha\beta}(\epsilon+j_0m^p_{\alpha})(Q^{-1})_{\alpha\beta}(\epsilon+j_0m^p_{\beta}) -\sum_{\alpha >\beta}q_{\alpha\beta}\bar q_{\alpha\beta} -\sum_{\alpha} m_{\alpha}\bar m_{\alpha}+\ln \Tr {\rm e}^L,
\end{eqnarray}
with
\begin{equation*}
Q_{\alpha\beta}=\left\{
\begin{array}{l l}
1 & (\alpha=\beta)\\
q_{\alpha\beta}^p & (\alpha\neq\beta)\\
\end{array}
\right.
\end{equation*}
and
\begin{equation*}
L=\sum_{\alpha > \beta}\bar q_{\alpha\beta}\sum_i S^{\alpha}_i S^{\beta}_i+\sum_{\alpha} \bar m_{\alpha}\sum_i S^{\alpha}_i,
\end{equation*}
where the sums over $\alpha$ and $\beta$ go from $1$ to $n$. Here, $q_{\alpha\beta}=N^{-1}\sum_i S_i^{\alpha}S^{\beta}$ is the spin glass order parameter and $m_{\alpha}=N^{-1}\sum_i S_i^{\alpha}$ is the ordinary magnetization. The symbols $\bar q_{\alpha\beta}$ and $\bar m_{\alpha}$ arise from technical operations and can be understood as the Fourier modes of $ q_{\alpha\beta}$ and $m_{\alpha}$, respectively. For details, see Appendix \ref{derivation}.

We will work mainly with the replica-symmetric ansatz, $q_{\alpha\beta}=q$ and $m_{\alpha}=m$, and consider replica symmetry breaking where necessary. As shown in Appendix \ref{sec:RSA}, the replica symmetric entropy is 
\begin{equation*}
s_{\textrm{RS}}=-\dfrac{1}{J^2}(\epsilon+j_0 m^p)^2\dfrac{1}{1-q^p}+\dfrac{1}{2}q\bar q-\dfrac{1}{2}\bar q-m\bar m +\int Du \ln 2\cosh(\sqrt{\bar q}u +\bar m),
\end{equation*}
where $Du=du \exp(-u^2/2)/(2\pi)^{1/2}$, with the saddle-point equations as given in equations (\ref{RS_SPE}a-d).

 To be able to compare the microcanonical phase diagram to the canonical one, we have to know the temperature of the microcanonical system. The inverse temperature is defined as the energy-derivative of the entropy,
\begin{equation}
\dfrac{1}{T}=\dfrac{\partial s_{\rm RS}}{\partial \epsilon}(m^*,q^*)=-\dfrac{2(\epsilon+j_0(m^*)^p)}{J^2(1-(q^*)^p)},
\end{equation}
where $m^*$ maximizes and $q^*$ minimizes the entropy simultaneously. Notice that a peculiarity of the replica trick is that while the equilibrium entropy has to be maximal with respect to $m$, it has to be minimal with respect to $q$.

\section{Phase Diagrams}\label{sec:comp}
In this section we compare the canonical and microcanonical phase diagrams of the model introduced in the previous section for finite $p$ and in the limit $p\rightarrow\infty$.

\subsection{Sherrington-Kirkpatrick model}
The Sherrington-Kirkpatrick model ($p=2$) \cite{SK} is well understood and has been studied extensively in the canonical ensemble. As only second-order phase transitions occur in this model, it is straightforward to confirm that the ensembles give equivalent results and lead to the same phase boundaries between the paramagnetic and ordered (ferromagnetic or spin glass) phases as long as the replica-symmetric solutions are concerned.

\subsection{Many-body interactions}
\begin{figure}[ht]
\centering
\includegraphics[width=\textwidth]{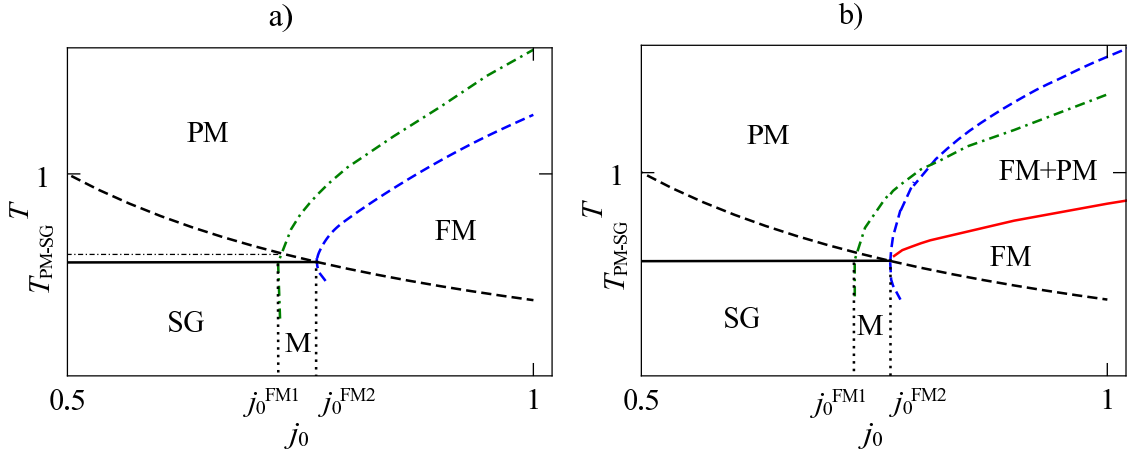} 
\caption{a) The canonical phase diagram of the model  with $p=3$. The equilibrium FM phase extends up to the blue dashed line and the spinodal line is drawn green dash-dotted, which marks the metastability limit of the FM phase. The horizontal thin dash-dotted line is the dynamical SG transition. b) The microcanonical phase diagram. Here, the FM phase extends up to the blue dashed line, while the PM solution extends down to the red line, resulting in a mixed phase FM+PM. The green dash-dotted line is the microcanonical spinodal line, which limits the metastability of the microcanonical FM phase. Note that the microcanonical spinodal line (green dash-dotted) can lie below the equilibrium FM boundary (blue dashed), because the energy, not the temperature, is the control parameter in the microcanonical ensemble. The specific heat can be negative, thus a higher temperature may mean lower energy. For details, see text and figure \ref{p3_mc_meta}. A dynamical SG-transition line as in a) may exist, but is omitted here due to lack of firm evidence. The black dashed line in both diagrams is the Nishimori line.} \label{p3_pd}
\end{figure}

The canonical phase diagram of the $p$-spin model with $p=3$ is also well known \cite{nishimori-wong,krzde2}. The paramagnetic (PM) and ferromagnetic (FM) phases are separated by a first order transition. The spin-glass phase (SG) appears only when we consider replica-symmetry breaking (discussed in Appendix \ref{derivation}) and the PM-SG transition is of second order. Between the FM and SG phases lies a mixed phase M where the FM solution shows replica symmetry breaking. The canonical phase diagram is depicted in figure  \ref{p3_pd} a). 

Figure \ref{p3_pd} b) shows the microcanonical phase diagram. The phase boundary between the PM and FM phases (red and blue dashed lines), as well as the spinodal line (green dash-dotted) were obtained by solving the replica symmetric saddle point equations (\ref{RS_SPE}a-d). The PM-SG phase boundary (black) was calculated from the first-step replica symmetry breaking equations (\ref{1RSB_SPE}a-e). In both ensembles the PM-SG transition occurs at $T_{\textrm{PM-SG}}=0.651$. The SG phase extends up to the critical value of the ferromagnetic bias of $j_0^{\textrm{FM2}}=0.768$, where we have chosen the unit $J=1$, and the metastability limit of the FM phase is at $j_0^{\textrm{FM1}}=0.736$. Between these two values of $j_0$ lies for low temperatures the region M with metastable FM solutions, while the SG and replica-symmetric FM phase meet directly around $T=T_{\textrm{PM-SG}}$ and $j_0=j_0^{\textrm{FM2}}$. Note that the replica-symmetry breakdown is used only for the SG solution in the present calculations.

\begin{figure}[ht]
\centering
\includegraphics[width=\textwidth]{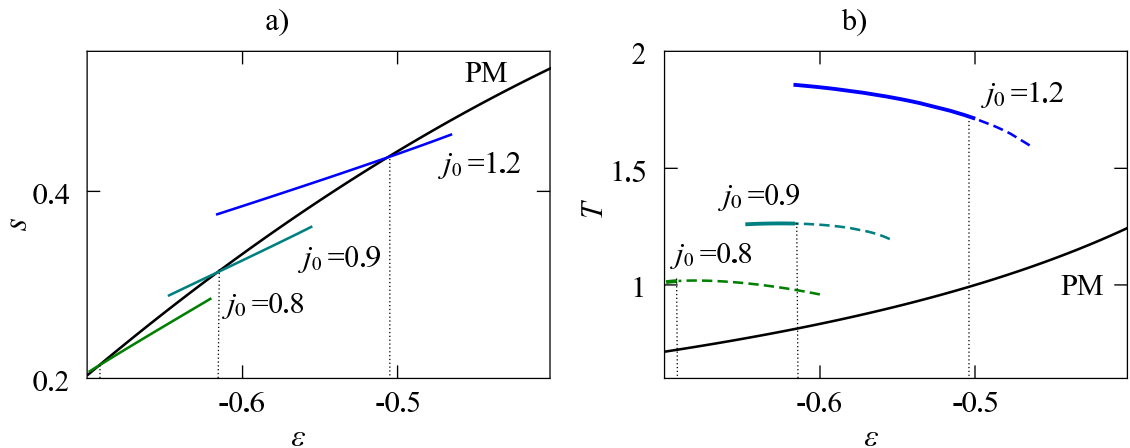} 
\caption{a) Microcanonical entropy versus energy per spin for various values of $j_0$, PM stands for the entropy of the paramagnetic solution. b) Corresponding caloric curves. The vertical dotted lines indicate the critical energies for $j_0=0.8$, $j_0=0.9$ and $j_0=1.2$. Negative specific heat occurs where $dT/d\epsilon<0$.  } \label{p3_mc_meta}
\end{figure}

The microcanonical phase diagram deviates from the canonical one only along the PM-FM transition. The PM phase in the microcanonical ensemble extends down to the red line where the PM solution ceases to be stable (to be called the $m=0$-line), while the FM phase extends up to the blue dashed line ($m>0$-line), resulting in a FM-PM mixed phase (FM+PM). We can understand the reason for the emergence of the FM+PM phase in the microcanonical ensemble when we look at the entropy as a function of the energy as depicted in figure \ref{p3_mc_meta} a). For each value of the bias $j_0$ there exists a FM solution to the saddle-point equations (\ref{RS_SPE}a-d) which crosses the PM entropy at some critical energy. Since the inclines of both curves are different, this results in a temperature jump at the transition as shown in \ref{p3_mc_meta} b). Thus, the $m=0$-line (red in figure \ref{p3_pd} b)) corresponds to the temperature of the PM solution at the transition, while the $m>0$-line (blue dashed in figure \ref{p3_pd} b)) corresponds to the FM solution.
The appearance of a `reentrant' tail of the FM transitions around the center of figure \ref{p3_pd} b) implies an instability of the replica-symmetric solution. In the canonical ensemble this is well understood through the de Almeida-Thouless line \cite{AT}, but the generalization to the microcanonical ensemble is quite involved and will not be developed here, mainly because it has no direct relevance for the ensemble inequivalence.

 We can also draw the Nishimori line (black dashed in figure \ref{p3_pd}), which was recently generalized to the microcanonical ensemble \cite{nishimori2}. The Nishimori line is a hypersurface in the parameter space of models with quenched randomness, where the averages of physical quantities take on simple values \cite{nishimori,nishimori0}. In the present case, the condition $T=1/(2j_0)$ defines the Nishimori line in the canonical ensemble and $\epsilon = - j_0$ in the microcanonical ensemble. An interesting feature is that on the Nishimori line there is no ensemble inequivalence \cite{nishimori2} and the microcanonical $m=0$- and $m>0$-lines meet there.

Part of the FM solution in the FM+PM phase for $j_0>0.92$ has negative specific heat closely below the critical energy. Below $j_0<0.92$ part of the metastable states have this property. To illustrate this point, we show the PM entropy  and short segments of the FM entropy for various values of $j_0$ in figure \ref{p3_mc_meta} a). The corresponding caloric curves are drawn in figure \ref{p3_mc_meta} b), where the metastable part is shown dashed, while the stable part is shown as a thicker line. Negative specific heat occurs when $dT/d\epsilon<0$. As $j_0$ decreases, so does the region of negative specific heat. Note that $dT/d\epsilon<0$ occurs only in the mixed region FM+PM and never in the pure FM phase.

The ergodicity is nowhere broken in this model. This is therefore an example for a spin system with infinite-range, many-body interactions in which not all features of ensemble inequivalence appear. 

The phase diagrams for $p=4$ and $p=5$ are qualitatively the same as for $p=3$, while the region M becomes narrower and $T_{\textrm{PM-SG}}$ becomes lower as $p$ grows. In table \ref{p_comp} we list the results of the critical values including results of the limit $p\rightarrow\infty$, discussed in the next section \ref{sec:rem}.

\begin{table}[ht]
{\begin{tabular}{@{}lccccc}\hline
 	& $p=2$ & $p=3$ & $p=4$ & $p=5$ & $p\rightarrow\infty$ \\
\hline

$T_{\textrm{PM-SG}}$ & 1  & 0.651 & 0.628 & 0.615 & $1/(2\sqrt{\ln 2})=0.601$\\
$j_0^{\textrm{FM1}}$ & 0.5 & 0.726 & 0.79 & 0.811 & $\sqrt{\ln 2}=0.833$ \\
$j_0^{\textrm{FM2}}$ & 0.5 & 0.767 & 0.81 & 0.82 & $\sqrt{\ln 2}=0.833$ \\
\hline
\end{tabular}}
\caption{Summary of critical values for various $p$ in the microcanonical ensemble.}
\label{p_comp}
\end{table}

\subsection{Random energy model}\label{sec:rem}

In the limit $p\rightarrow\infty$, the model \eqref{HAM} is known as the random energy model \cite{derrida}. The energy levels are Gaussian-distributed random variables. It is also known to be the simplest spin glass \cite{gross} and its canonical phase diagram can be obtained analytically in a simple, straightforward manner. The SG phase in the canonical ensemble does not show up with the replica-symmetric ansatz, but is fully described by the first-step replica symmetry breaking (1RSB). The FM and PM phases are replica symmetric.

Let us first derive the PM-FM phase boundary in the microcanonical ensemble. Here, we can still use the replica-symmetric ansatz. In the limit $p\rightarrow\infty$, the saddle-point equations (\ref{RS_SPE}a-d) have the following solutions: $q=m=0$ (PM) and $q=m=1$ (FM), the latter being together with the requirement $\epsilon=-j_0$.
Taking the same limit, $p\rightarrow\infty$, for the 1RSB saddle-point equations (\ref{1RSB_SPE}a-e) it is straightforward to see, as in the canonical case, that the only other relevant solution, apart from the replica symmetric FM and PM solutions, is the SG solution $q_1=1$, $q_0=0$ and $m=0$. To obtain the phase diagram, we list the entropies of the different phases,
\begin{align}
& s_{\textrm{PM}} (\epsilon) = \ln 2 - \epsilon^2 \quad( -\sqrt{\ln 2}\leq \epsilon \leq 0) \\
& s_{\textrm{FM}} (\epsilon) = 0 \quad(\epsilon =-j_0)\\
& s_{\textrm{SG}} (\epsilon) = \frac{1}{x}(\ln 2 - \epsilon^2)\quad( -\sqrt{\ln 2}\leq \epsilon \leq 0),
\end{align}
where $x(0\leq x \leq1)$ is the 1RSB parameter, the boundary between $q_0$ and $q_1$ in the 1RSB matrix (see Appendix \ref{sec:1RSB} for details). In the energy range $-\sqrt{\ln 2} \leq \epsilon\leq 0$, the two entropies $s_{\textrm{PM}} $ and $s_{\textrm{SG}} $ compete but the former wins because $s_{\textrm{PM}} < s_{\textrm{SG}} $ as long as $0 \leq x < 1$. As mentioned before, as far as the SG phase is concerned, the entropy has to be minimal. 

\begin{figure}[htb]
\centering
\includegraphics[width=\textwidth]{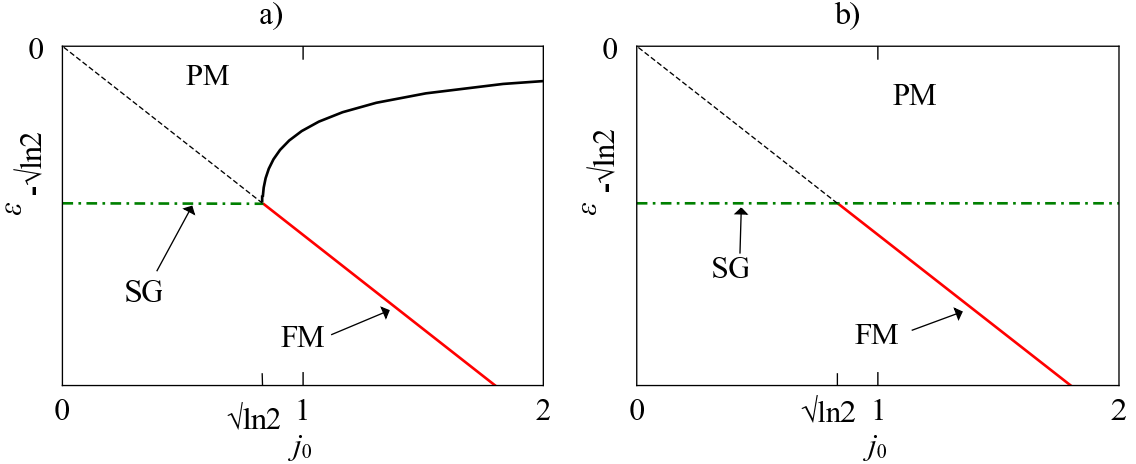} 
\caption{ a) The canonical phase diagram of the random energy model with the energy $\epsilon$ and the ferromagnetic bias $j_0$ as axes. The PM phase exists at energies above $\epsilon\geq-\sqrt{\ln 2}$ for $j_0<\sqrt{\ln 2}$ and above the black line for $j_0\geq\sqrt{\ln 2}$. The SG phase exists at $\epsilon=-\sqrt{\ln 2}$ for $j_0\leq\sqrt{\ln 2}$, shown green dash-dotted while the FM phase (red) has the energy $\epsilon=-j_0$ for $j_0>\sqrt{\ln 2}$. b) The microcanonical phase diagram. The PM phase exists for energies $\epsilon\geq-\sqrt{\ln 2}$, while the SG phase manifests itself only at the energy $\epsilon=-\sqrt{\ln 2}$ (green dash-dotted) for all values of $j_0$. The FM phase (red) shows the same behavior as in the canonical ensemble, it has $\epsilon=-j_0$ for $j_0>\sqrt{\ln 2}$ and does not show up for a smaller bias, $j_0<\sqrt{\ln 2}$. The black dashed line in both diagrams is the Nishimori line, which coincides with the ferromagnetic solution (red) for $j_0>\sqrt{\ln 2}$} \label{pinf_pd_comp}
\end{figure}

When $x = 1$, the two phases merge, but, on the other hand, $x = 1$ implies replica symmetry. Thus, there is no SG phase except at $\epsilon = -\sqrt{ \ln 2}$, where the two phases merge, since both have zero entropy. The FM phase is allowed only when the energy takes the specific value $\epsilon= -j_0$ .

For any value of $j_0$ the FM and SG phases are defined only at a single energy respectively, and therefore it is not possible to define a temperature for these phases. We cannot draw a standard phase diagram with the temperature and $j_0$ as axes to compare the microcanonical to the canonical phase diagram. We can, however, draw a canonical phase diagram with the energy and $j_0$ as the axes and thus compare the ensembles. This is done in figure \ref{pinf_pd_comp}. We see that the phase diagrams are almost equivalent, except that in the microcanonical case, figure \ref{pinf_pd_comp} b), the PM extends down to $\epsilon=-\sqrt{\ln 2}$ for any $j_0$. 

\begin{figure}[ht]
\centering
\includegraphics[width=0.5\textwidth]{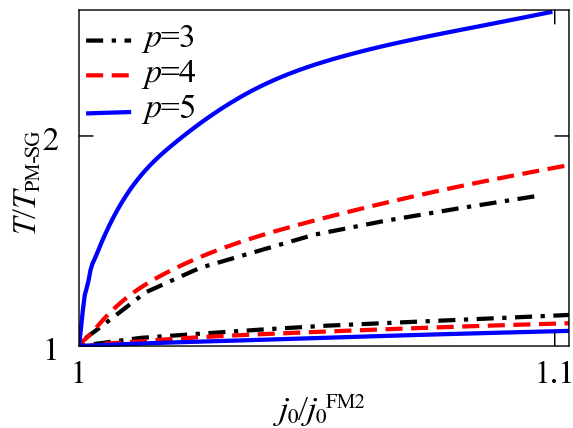} 
\caption{The microcanonical curves show that the mixed region FM+PM, which lies between the two curves of the respective $p$, extends in an ever wider region for growing $p$.} \label{mc_comp}
\end{figure}

\begin{figure}[ht]
\centering
\includegraphics[width=\textwidth]{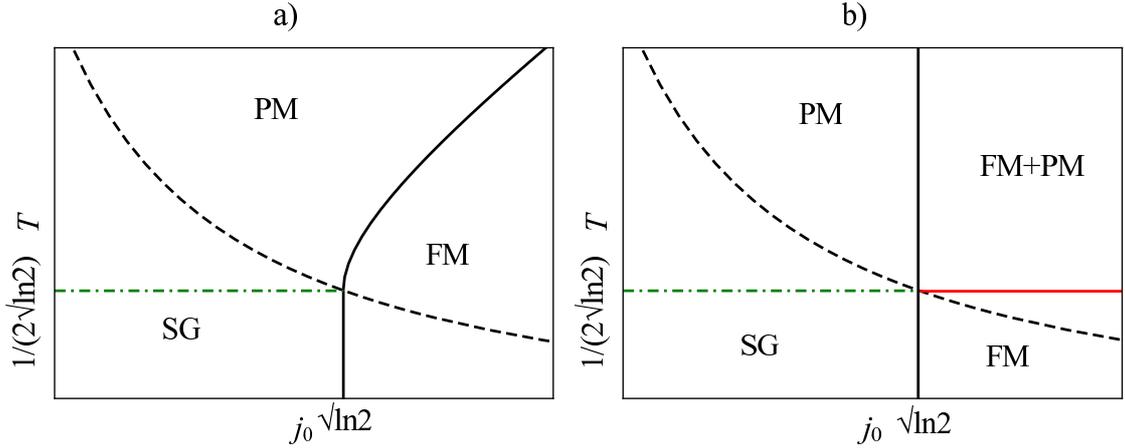} 
\caption{a) Canonical phase diagram of the random energy model. b) Microcanonical phase diagram as inferred from the limit of large $p$. The dashed line in both diagrams is the Nishimori line.} \label{pinf_pd_comp_b}
\end{figure}

For a clearer understanding, we investigate how the microcanonical phase diagram evolves as $p$ grows larger. To that end, we draw in figure \ref{mc_comp} the microcanonical phase boundaries between the PM and FM phases, scaled to their respective critical values (see table \ref{p_comp}) for $p=3$ (black dash-dotted), $p=4$ (red dashed) and $p=5$ (blue). As the curves suggest, the area of the mixed phase FM+PM grows larger with $p$. We can infer that in the limit of very large $p$ the pure FM phase extends only to the temperature where the PM phase freezes out for $j_0<j_0^{\rm FM2}$. The spinodal line and the $m>0$-line coincide in the limit $p\rightarrow\infty$ and the mixed phase M, of the FM and SG states, will disappear in this limit. The FM+PM phase extends over the whole quarter-plane right to the critical $j_0=\sqrt{\ln 2}$ and above $T=1/(2\sqrt{\ln2})$. The SG phase is the same as in the canonical case, while the pure PM phase exists above $T=1/(2\sqrt{\ln 2})$ and left of  $j_0=\sqrt{\ln 2}$ as shown in figure \ref{pinf_pd_comp_b} b). However, since neither the SG nor the FM phase have clearly defined temperatures when $p=\infty$, the re-interpretation of figure \ref{pinf_pd_comp} b) into the ($T,j_0$) plane as drawn in figure \ref{pinf_pd_comp_b} b) is to be taken with caution and has to be contrasted to results that were obtained in the microcanonical ensemble without the use of the replica trick \cite{derrida,gross,OCY}. 

\section{Conclusion}\label{sec:summary}
We have derived the microcanonical entropy of the $p$-spin model with Gaussian-distributed random interactions with the help of the replica trick. To the best of our knowledge, our study represents the first consequent treatment of spin glasses in the microcanonical ensemble.
For $p=2$, we recover the results of Sherrington and Kirkpatrick, which was to be expected, as there are only second-order phase transitions present. For $p>2$, many-body interactions effect a first-order transition between the PM and FM phases and part of the FM solution in the mixed phase FM+PM has negative specific heat above a certain value of $j_0$, but never in the pure FM phase. This system does not show ergodicity breaking, which means that all metastable states have finite lifetimes for finite systems. We have also taken the limit $p\rightarrow\infty$ within the framework of the replica theory and could show that the two ensembles give different results for the random energy model.   

Our study shows that great care has to be taken in the analysis of many-body spin glasses, and even in well-understood systems new properties can emerge. This fact should particularly be taken seriously in the spin-glass context, where the realistic finite-dimensional properties are often inferred from the analysis of mean-field limits.

\section*{Acknowledgements}
We thank Kazutaka Takahashi for fruitful discussions and useful comments.

\appendix
\section{Derivation of the Entropy}\label{derivation}
In this section we will derive the entropy from the replicated number of states in detail and calculate the form of the entropy in the replica-symmetric and the first-step replica-symmetry breaking ansatz.

The Hamiltonian is given by
\begin{equation*}
H=-\sum _{i_1<...<i_p}J_{i_1...i_p}S_{i_1}...S_{i_p},
\end{equation*}
where the interactions are distributed via
\begin{equation*}
P(J_{i_1...i_p})=\left(\dfrac{N^{p-1}}{\pi J^2 p!}\right)^{1/2}\exp \left(-{\left(J_{i_1...i_p}-\dfrac{j_0p!}{N^{p-1}}\right)^2\dfrac{N^{p-1}}{J^2 p!}}\right)\equiv \sqrt{\dfrac{F}{\pi}}{\rm e}^{-(J_{i_1...i_p}-j_0/F)^2 F}.
\end{equation*}
We want to calculate the microcanonical entropy per spin: $s=\ln\Omega/N$ where
\begin{equation}
\Omega = \Tr \delta (E-H)=\Tr\int \dfrac{dk}{2\pi} {\rm e}^{{\rm i}k(E-H)},
\end{equation}
and the trace is over the Ising spins.

We use the replica trick and can closely follow the calculations for the canonical ensemble \cite{nishimori}:
%\begin{equation}
$[\ln\Omega]=\lim_{n\rightarrow 0}{([\Omega^n] -1)}/{n}$.
%\end{equation}
The replicated sum of states reads 
\begin{multline}
[\Omega^n]=\\
\int \prod_{i_1<..<i_p}dJ_{i_1...i_p}\sqrt{\dfrac{F}{\pi}}{\rm e}^{-(J_{i_1...i_p}-j_0)^2 F} \Tr\int \left(\prod_{\alpha}\dfrac{dk_{\alpha}}{2\pi}\right) {\rm e}^{{\rm i}\sum_{\alpha} k_{\alpha}(E+\sum J_{i_1...i_p} S^{\alpha}_{i_1}...S^{\alpha}_{i_p})},
\end{multline}
where $\alpha$ runs from $1$ to $n$.
We write $\left(\prod_{\alpha}{dk_{\alpha}}/{2\pi}\right)=Dk$ and integrate over the $J_{i_1..i_p}$ to get
\begin{multline}
[\Omega^n]= \Tr \int Dk \exp\left({{\rm i}E\sum_{\alpha} k_{\alpha}-\dfrac{1}{4F}\sum_{\alpha,\beta}k_{\alpha}k_{\beta}\sum_{i_1<...<i_p}S^{\alpha}_{i_1}...S_{i_p}^{\alpha}S_{i_1}^{\beta}...S_{i_p}^{\beta}} \right)\\
\times\exp\left({\rm i}\dfrac{j_0}{F}\sum_{\alpha} k_{\alpha}\sum_{i_1<...<i_p}S^{\alpha}_{i_1}...S_{i_p}^{\alpha}\right).
\end{multline}
Now, we make use of the relation
\begin{equation}
\dfrac{1}{N^{p-1}}\sum_{i_1<...<i_p}S_{i_1}...S_{i_p}\approx\dfrac{N}{p!}\left(\dfrac{1}{N}\sum_i S_i\right)^p,
\end{equation}
which holds for $N\gg1$. We set $J=1$ for simplicity and have
\begin{multline}
[\Omega^n]= \Tr \int Dk \exp\left\{ N {\rm i}\sum_{\alpha} k_{\alpha}\left(E/N+j_0\left[\dfrac{1}{N}\sum_{i}S^{\alpha}_{i}\right]^p\right)\right\}  \\
 \times\exp{\left(-\dfrac{N}{4}\sum_{\alpha,\beta}k_{\alpha}k_{\beta}\left[\dfrac{1}{N}\sum_{i}S^{\alpha}_{i}S_{i}^{\beta}\right]^p\right)} \nonumber. 
\end{multline}
It is convenient to introduce $\delta$-functions $\delta(m_{\alpha}-1/N\sum_iS_i^{\alpha})$ and  $\delta(q_{\alpha\beta}-1/N\sum_iS_i^{\alpha}S_i^{\beta})$ for $\alpha>\beta$. Then by rewriting $\epsilon_{\alpha}=E/N+jm^p$ and performing the integrals over $k$, we arrive at 
%\begin{multline}
%[\Omega^n]= \Tr \int Dk \ Dm \ Dq \exp\left({Ni\sum_{\alpha} k_{\alpha}\epsilon_{\alpha}-N\sum_{\alpha,\beta}k_{\alpha}Q_{\alpha\beta} k_{\beta}} \right)\\
%\times\prod_{\alpha}\delta(m_{\alpha}-1/N\sum_iS_i^{\alpha})\prod_{\alpha>\beta}\delta(q_{\alpha\beta}-1/N\sum_iS_i^{\alpha}S_i^{\beta}).
%\end{multline}
% with $Dq=\prod dq$ and $Dm=\prod dm$. Here, $Q_{\alpha\beta}=1$ for $\alpha=\beta$ and $Q_{\alpha\beta}=q_{\alpha\beta}^p$ otherwise.
% We can now perform the integral over $k$ and get 
\begin{multline}
[\Omega^n]\propto  \Tr \int Dm \ Dq \exp\left(-{N}\sum_{\alpha\beta}\epsilon_{\alpha}(Q^{-1})_{\alpha\beta} \epsilon_{\beta}\right)\\
\times\prod_{\alpha}\delta(m_{\alpha}-\dfrac{1}{N}\sum_iS_i^{\alpha})\prod_{\alpha>\beta}\delta(q_{\alpha\beta}-\dfrac{1}{N} \sum_iS_i^{\alpha}S_i^{\beta}),
\end{multline}
with $Dq=\prod dq$, $Dm=\prod dm$ and $Q_{\alpha\beta}=1$ for $\alpha=\beta$ and $Q_{\alpha\beta}=q_{\alpha\beta}^p$ otherwise.
Inserting the integral representations of the $\delta$-functions 
%\begin{equation}
%\delta(x-y)\propto \int_{-i\infty}^{i\infty}e^{-N(x-y)k}dk
%\end{equation}
and collecting the terms proportional to $N$ allows us to evaluate the integral with the saddle point method as
\begin{eqnarray}\label{OMEGA}
[\Omega^n]\propto \exp N \left(-\sum_{\alpha\beta}\epsilon_{\alpha}(Q^{-1})_{\alpha\beta}\epsilon_{\beta} -\sum_{\alpha >\beta}q_{\alpha\beta}\bar q_{\alpha\beta} -\sum_{\alpha} m_{\alpha}\bar m_{\alpha}+\ln \Tr {\rm e}^L\right)
\end{eqnarray}
with
\begin{equation}
L=\sum_{\alpha > \beta}\bar q_{\alpha\beta}\sum_i S^{\alpha}_i S^{\beta}_i+\sum_{\alpha} \bar m_{\alpha}\sum_i S^{\alpha}_i.
\end{equation}
\subsection{Replica-symmetric solution}\label{sec:RSA}
If we set $q_{\alpha\beta}=q$ and $m_{\alpha}=m$, we have
\begin{equation}
Q_{\alpha\beta}=\left\{
\begin{array}{l l}
1 & (\alpha=\beta)  \\
q^p & (\alpha\neq\beta) \\
\end{array}
\right. .
\end{equation}
Then the sum over the elements of the inverse of $Q_{\alpha\beta}$ is calculated by a very simple trick:
\begin{eqnarray*}
&&\sum_{\gamma}Q_{\alpha\gamma}(Q^{-1})_{\gamma\beta}=\delta_{\alpha\beta}\\
&&\sum_{\alpha\gamma}Q_{\alpha\gamma}\sum_{\beta}(Q^{-1})_{\gamma\beta}=n.
\end{eqnarray*}
Noting that the sum of all columns of $Q_{\alpha\beta}$ yields the same result, $\sum_{\beta}Q_{\alpha\beta}=1+(n-1)q^p$, we arrive at
\begin{equation}
\sum_{\alpha\beta}(Q^{-1})_{\alpha\beta}=\dfrac{n}{1+(n-1)q^p}.
\end{equation}
We  next have to trace over the Ising spins. Using
\begin{equation}
L=\bar q\left(\sum_{\alpha} S^{\alpha}\right)^2+\bar m\sum_{\alpha}  S^{\alpha} -\dfrac{n}{2}\bar q,
\end{equation}
we can take the $n\rightarrow 0$ limit as
\begin{equation}
\lim_{n\rightarrow 0}\dfrac{1}{n}\ln \Tr {\rm e}^L=\int Du \ln 2\cosh(\sqrt{\bar q}u +\bar m)-\dfrac{1}{2}\bar q,
\end{equation}
with $Du=\exp({-u^2/2})/\sqrt{2\pi}$. 

The entropy defined by $s=[\Omega^n]/nN$ now becomes in the limit $n \to 0,N\to \infty$ 
\begin{equation}\label{RS_p_entropy}
s_{\textrm{RS}}=-\dfrac{1}{J^2}(\epsilon+j_0 m^p)^2\dfrac{1}{1-q^p}+\dfrac{1}{2}q\bar q-\dfrac{1}{2}\bar q-m\bar m +\int Du \ln 2\cosh(\sqrt{\bar q}u +\bar m),
\end{equation}
with the saddle point equations
\begin{subequations}\label{RS_SPE}
\begin{align}
& q=\int Du\tanh^2(\sqrt{\bar q}u+\bar m)\\
& m=\int Du\tanh(\sqrt{\bar q}u+\bar m)\\
& \bar q=\dfrac{2}{J^2}(\epsilon+j_0 m^p)^2\dfrac{pq^{p-1}}{(1-q^p)^2}\\
& \bar m=-\dfrac{2j_0}{J^2}(\epsilon+j_0 m^p)\dfrac{pm^{p-1}}{1-q^p}.
\end{align}
\end{subequations}

\subsection{First-step Replica Symmetry Breaking}\label{sec:1RSB}
The first-step replica symmetry breaking is characterized by splitting the diagonal of the matrix $q_{\alpha\beta}$ into $x$ $\times$ $x$ blocks and filling the off-diagonal elements of these blocks with $q_1$ and the rest of the matrix with $q_0$, while the diagonal is filled with $0$.

As such, the entropy is calculated from (\ref{OMEGA}) as $s=\ln [\Omega^n]/nN$. The corresponding terms are calculated as
\begin{equation}
\dfrac{1}{n}\sum_{\alpha\beta}(Q^{-1})_{\alpha\beta}=\dfrac{1}{1+(x-1)q_1^p+(n-x)q_0^p}=\dfrac{1}{1-(1-x)q_1^p-xq_0^p},
\end{equation}
\begin{equation}
\dfrac{1}{n}\sum_{\alpha\beta} \bar q_{\alpha\beta}q_{\alpha\beta}=n\bar q_0 q_0 +x^2(\bar q_1q_1-\bar q_0q_0)\dfrac{1}{x}-\bar q_1q_1=-(1-x)\bar q_1 q_1-x\bar q_0q_0
\end{equation}
and
\begin{equation}
\dfrac{1}{n}\ln \Tr {\rm e}^L=\dfrac{1}{x}\int Du\ln \int Dv \cosh^x\left(u\sqrt{\bar q_0}+v\sqrt{\bar q_1-\bar q_0}+\bar m\right)-\dfrac{1}{2}\bar q_1+\ln 2,
\end{equation}
where $Dv$ is defined in the same way as $Du$.
The entropy is
\begin{multline}
s_{\textrm{1RSB}} =\ln 2-\dfrac{(\epsilon+j_0m^p)^2}{J^2(1-(1-x)q_1^p-xq_0^p)}+\dfrac{1}{2}((1-x)\bar q_1 q_1-x\bar q_0q_0-\bar q_1)-\bar m m\\
 +\dfrac{1}{x}\int Du\ln \int Dv \cosh^x\left(u\sqrt{\bar q_0}+v\sqrt{\bar q_1-\bar q_0}+\bar m\right)
\end{multline}
and the saddle point equations are
\begin{subequations}\label{1RSB_SPE}
\begin{align}
&\bar q_a=\dfrac{(\epsilon+j_0m^p)^2 2pq_a^{p-1}}{J^2(1-(1-x)q_1^p-xq_0^p)^2}\quad\quad (a=0,1)\\
&\bar m=-\dfrac{2pj_0m^{p-1}(\epsilon+j_0m^p)}{J^2(1-(1-x)q_1^p-xq_0^p)}\\
& q_0=\int Du \left(\dfrac{\int D'v \tanh K }{\int D'v}\right)^2\\
& q_1=\int Du \dfrac{\int D'v \tanh^2 K }{\int D'v}\\
& m=\int Du \dfrac{\int D'v \tanh K }{\int D'v}\\
& K=u\sqrt{\bar q_0}+v\sqrt{\bar q_1-\bar q_0}+\bar m\\
& D'v=dv \exp(-v^2/2)\cosh^x K.
\end{align}
\end{subequations}

\end{document}